\documentclass[12pt]{iopart}
\usepackage{graphicx}
\usepackage{iopams}
\begin{document}

\title[Bragg Birefringence]{Birefringence in periodic holographic grating}
\author{Gagik T Avanesyan$^{1,2}$}
\address{$^1$Institute of Mathematics, NAS - Yerevan, 0019, Armenia}
\address{$^2$Institute for Physical Research, NAS - Ashtarak, 0203, Armenia}
\ead{avangt@instmath.sci.am}
\begin{abstract}

Photonic bandgap in holographic grating manifests itself as phase-sensitive birefringence. Phenomenological theory of the effect is presented.

\end{abstract}


\section{Introduction}
Microscopic mechanisms behind the formation of holographic grating are complicated. Respective nonlinear system of differential equations for the kinetics of the process, which includes many parameters that are used to characterize the photorefractive material, can be solved through an intricate numerical calculation only, along with making a series of simplifying assumptions. Therefore, it is interesting to construct simpler phenomenological models to cope with the problem.

A simpler approach can be developed towards the non-equilibrium steady states (NESS) that are reached when the diffraction efficiency of the grating changes no more under stable irradiation \cite{Collier}. This is a NESS that will be characterized here by a single mode of the electromagnetic field (the writing mode), and the respective modulation of the electric permittivity (the photorefractive response). The two latter quantities are interdependent, and should be obtained in a self-consistent manner in the theory. (One may think of some functional, which has extrema at NESS "points" of the system driven by irradiation.)

From the viewpoint of nonequilibrium thermodynamics, the driven open system that is represented here by the electromagnetic field coupled to the photorefractive medium, slowly evolves towards a NESS under fast external driving.  The two time-scales differ in a dozen of orders in magnitude, and it is a natural phenomenological assumption that the modulation of the permittivity of the initially homogeneous medium is determined by the NESS mode itself. Based on this assumption is the discussion below.

An interesting feature of locally isotropic photorefractive medium - phase-sensitive birefringence effect, that is a manifest of the photonic bandgap and Bragg reflection, shows up.

\section{Exactly solvable model}\label{solve}
Vector nature of light is at the heart of the photorefractive effect. Fortunately, the two polarization modes can be treated separately, in the properly symmetric arrangements. In view of the holographic grating recording, of major interest are the modes of the interference pattern. Note that their observation can be perfectly "tuned in" due to Bragg reflection. By assuming that the active NESS mode is the electric component of the so-called E-wave,
\begin{equation}\label{ewave}
{E_y}(z,x,t)={F}\left(z\right){sin}({k_{x}} x-\omega t),
\end{equation}
Maxwell's equations are reduced to the second-order differential equation for the mode profile, $F(z)$:
\begin{equation}\label{maxwell}
    {F''}+(k_0^2{\varepsilon} (z)-k_x^2){F}=0;
\end{equation}
for details see \cite{LLVIII}. $F(z)$ is referred to below as the "NESS mode".
It is assumed that the in-medium mode inherits i) the antiperiodicity of the driving field, $F(z+\lambda_z/2)=-F(z)$, and, similarly, ii) the reflection symmetry for the mode profile, $F(-z)=F(z)$, or $F(-z)=-F(z)$, and, consequently, the eq. (\ref{maxwell}) has to be solved under antiperiodic boundary condition for the even or the odd mode.

Perhaps the simplest model assumption for the NESS permittivity is that that its modulation follows the intensity of the NESS mode directly; then the transparent and nonmagnetic medium of the model, with periodic inhomogeneity in the $z$-direction, is described by its permittivity at a given frequency $k_0$ of the mode by
\begin{equation}\label{epsilon1}
{\varepsilon}(z)={\epsilon}-2{\eta}F(z)^2+{\eta}F_0^2;
\end{equation}
$F_0$ is the amplitude of the NESS mode, and it is assumed that the ${\eta}F_0^2$ is relatively small number: the modulation $-{\eta}F_0^2$$\leq$${\varepsilon}(z)-\epsilon$$\leq$${\eta}F_0^2$ , typically, is weak in holographic gratings. An expression similar to (\ref{epsilon1}) is used in nonlinear electrodynamics to study self-defocusing of the electromagnetic waves (see \cite{LLVIII}). Here its meaning is different: we consider the linear electrodynamics problem, and the "nonlinear" term describes the mode profile in the NESS. In view of the localization properties of electrons in photorefractive materials, which are known to be good insulators, the local approximation in (\ref{epsilon1}) seems to be reasonable: it is a natural phenomenological assumption that the modulation of the permittivity of the initially homogeneous medium, in the first approximation, is a quadratic function of the NESS mode amplitude.

\begin{figure}
  \includegraphics[width=3 in]{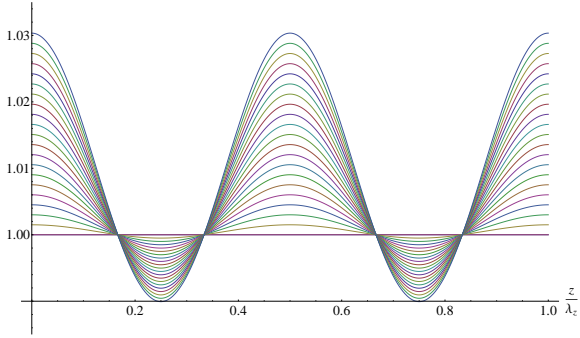}\\
  \caption{Permittivity modulations for $0\leq m \leq 0.02$. Plotted is the right-hand side of the eq. (\ref{prof}).}\label{lame}
\end{figure}

After substituting $\varepsilon(z)$ as a function of the "brightness" of the mode,  ${\varepsilon}(z)\rightarrow{\epsilon}-2{\eta}F^2+{\eta}F_0^2$, the linear equation (\ref{maxwell}) takes its nonlinear form:
\begin{eqnarray}
\label{maxwellness}
    {F}''+(k_0^2{\epsilon}-k_x^2+{\eta}k_0^2 F_0^2-2{\eta}k_0^2 F^2){F}=0.\nonumber
\end{eqnarray}
A first integral can be written down explicitly:
\begin{eqnarray}
\label{first}
  (F')^2+(k_0^2{\epsilon}-k_x^2+{\eta}k_0^2 F_0^2){F^2}-{\eta}k_0^2 F^4=(k_0^2{\epsilon}-k_x^2){F_0^2}.\nonumber
\end{eqnarray}
The wavelength of the mode is readily found:
\begin{eqnarray}
\label{firstg}
 \frac{{\lambda}_z }{2}= \int_{-F_0}^{F_0} \frac{1}{\sqrt{(F_0^2-F^2)(k_0^2{\epsilon}-k_x^2-{\eta}k_0^2 F^2)}} \, {dF};\nonumber
\end{eqnarray}
see textbooks on classical mechanics. The mode of interest is, naturally, that of winding number 1. Resulting dispersion relation has the form:
\begin{eqnarray}
\label{drelation}
 {\lambda}_z^2 (k_0^2{\epsilon}-k_x^2)=(4K(m))^2=\nonumber\\
(2\pi)^2(1+\frac{m}{2}+\frac{11 m^2}{32}+O\left(m^3\right)),
\end{eqnarray}
in contrast with the relation ${\lambda}_z^2 (k_0^2{\epsilon}-k_x^2)=(2\pi)^2$ for an isotropic and uniform medium. (We have avoided using the notation "$k_z$" along with the "$k_x$" to stress the fact that the former is actually a quasimomentum.) $K(m)$ is the complete elliptic integral of the first kind, and its modulus, $m$, is related to the parameters of the problem by the following equation:
\begin{eqnarray}
\label{modulus}
m=\eta  F_0^2\frac{ k_0^2}{k_0^2 \epsilon-k_x^2}.\nonumber
\end{eqnarray}
The NESS mode has the form
\begin{eqnarray}
\label{sn}
F(z)=F_0{sn}\left(\left.\frac{4 K(m) z}{\lambda_z }\right|m\right)\nonumber
\end{eqnarray}
with the Jacobi elliptic function $sn$. The respective profiles of the permittivity modulation normalized to the effective NESS mode,
\begin{eqnarray}\label{prof}
 \frac {{\lambda}_z^2 (k_0^2{\varepsilon} (z)-{k_x}^2)}{(2\pi)^2}=\frac{(4K(m))^2}{(2\pi)^2}\times\nonumber\\
\left(1-2m\left({sn}\left(\left.\frac{4 K(m) z}{\lambda_z }\right|m\right)\right)^2+m\right),
\end{eqnarray}
are shown in the \fref{lame}.

The eq. (\ref{maxwell}) with $\varepsilon(z)$ defined using (\ref{sn}) reduces to the well known equation of Lam\'{e}. With the brightest point of the interference pattern at $z=0$ the Jacobi $sn$ in (\ref{maxwellness}) and (\ref{sn}) will be replaced by the Jacobi $cd$. For references on Jacobi elliptic functions and Lam\'{e} equation see \cite{WW}.

Remarkably, the inverted brightness $(F^*(z))^2=F_0^2-(F(z))^2$ mode also exists as an eigenmode for the eq. (\ref{maxwell}),
\begin{eqnarray}
\label{cn}
F^*(z)=F_0{cn}\left(\left.\frac{4 K(m) z}{\lambda_z }\right|m\right);\nonumber
\end{eqnarray}
the ${cn}$ will be replaced by the $\sqrt{1-m}{sd}$ when the brightest point of the interference pattern is at $z=0$. The dispersion relation for these latter modes has the form
\begin{eqnarray}\label{drelationinv}
{\lambda}_z^2 (k_0^2{\epsilon}-{k_x^*}^2)=(1-m)(4K(m))^2=\nonumber\\
(2\pi)^2(1-\frac{m}{2}-\frac{5 m^2}{32}+O\left(m^3\right)),
\end{eqnarray}
and, therefore, the difference
\begin{eqnarray}
\label{birefrin}
{\lambda}_z^2 ({k_x^*}^2-k_x^2)=m(4K(m))^2=\nonumber\\
(2\pi)^2(m+\frac{m^2}{2}+O\left(m^3\right))
\end{eqnarray}
is small as expected. The picture is sensitive to the relative phase, and the dispersion relations (\ref{drelation}) and (\ref{drelationinv}) can be contrasted with those for the conventional birefringence case,
\begin{eqnarray}
\label{birefrino}
{\lambda}_z^2 ({k_0^2\varepsilon}_o-{k_{xo}}^2)=(2\pi)^2,{\lambda}_z^2 ({k_0^2\varepsilon}_e- {k_{xe}}^2)=(2\pi)^2\frac{{\varepsilon}_e}{{\varepsilon}_o},\nonumber
\end{eqnarray}
with the indices $o$ and $e$ denoting the ordinary and extraordinary waves, respectively.

The two solutions found belong to the band edges of the spectrum of the Hamiltonian associated with the periodic modulation of the $\varepsilon(z)$; the bandgap here manifests itself as a phenomenon resembling birefringence: the above expression (\ref{birefrin}) describes \emph{phase-dependent} birefringence; joint dispersion relation in parametric form is given by (\ref{drelation}) and (\ref{drelationinv}), and the small parameter $m$ is accounted for by (\ref{birefrin}). With the latter equation, experimental observation is straightforward, by measuring the difference between $k_x^*$ and $k_x$. Regarding such an experiment, as well as the symmetries of the problem, the remarks in the following section seem relevant.

\section{NESS, two basic geometries, and translation invariance}\label{ness}
\begin{figure}
  \includegraphics[width=3 in]{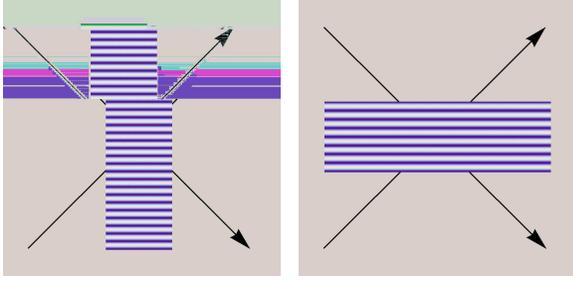}\\
  \caption{Two complementary setups for holographic grating recording. The two incident plane waves dictate either $\lambda_z$ (on the left) or $k_x$ (on the right). }\label{setups}
\end{figure}

As regards the (relative phase-sensitive) "Bragg birefringence" effect mentioned at the end of the previous section, two setups shown in the \fref{setups} have to be considered. Full translation invariance in the bulk of the photorefractive medium is conserved along the horizontal axis only (the $x$-axis). With the boundaries of the plate parallel to the vertical axis (the $z$-axis), externally controlled is the quantity $\lambda_z$, and the quantity $k_x$ - in the next setup, with the boundaries parallel to the $x$-axis. Also notice that the parameter $\epsilon$ that enters the joint dispersion relation (\ref{drelation}) and (\ref{drelationinv}) is "renormalized" by the NESS grating, and actually also depends on the $m$, as a self-consistent quantity. Therefore, to use the dispersion relations, the small parameter $m$ has to be determined first, by using the relation (\ref{birefrin}).
This can be done, in this example, due to the reflection symmetry: by externally exciting the two modes, $F$ and $F^*$, separately.

Interference pattern created by two crossing plane waves in a homogeneous transparent medium - the intensity of the superposition mode of the electromagnetic field, is considered to be the "driving" field. For instance, the superposition mode (the two incident waves have opposite phases at the origin of the coordinate system)
\begin{eqnarray}
\label{trigs}
{E_y}\sin ({k_z} z) \sin ({k_x} x-\omega t)\nonumber
\end{eqnarray}
creates a one-dimensional interference pattern with intensity profile $\frac{1}{2}({E_y}\sin({k_z} z))^2$, and, similarly, the intensity profile is $\frac{1}{2}({E_y}\cos ({k_z} z))^2$ for the mode
 \begin{eqnarray}
 \label{trigc}
{E_y}\cos ({k_z} z) \cos ({k_x} x-\omega t)\nonumber
\end{eqnarray}
(for coinciding at the origin phases). As regards the interpretation of the (\ref{epsilon1}) in view of nonequilibrium processes for the E-wave, the entropy production in transparent materials is  actually the Joule heat, and the modified $E_y^2$ is at place in the equation.

The period of the interference pattern is ${\lambda_z}/{2}$, and the bright and the dark strips alternate at half that length.
Translation by ${\lambda_z}/{4}$,
\begin{equation}\label{shifts}
    F(z)\rightarrow \tilde{F}(z)\equiv F\left(z+{\lambda_z}/{4}\right),
\end{equation}
results in swapping them (recall that ${\lambda_z}/{2}$-antiperiodic modes are considered, and, therefore, the direction of the shift could be reversed as well). For the brightness of the interference pattern in a homogeneous medium, the following symmetry property holds:
\begin{equation}\label{symmetry}
    {\tilde{F}(z)}^2={{{F}}_0}^2-{{F}}(z)^2.
\end{equation}
For the NESS, however, full translation invariance of the medium is broken due to the electric permittivity modulation: the "polaritonic" nature of the NESS mode should not be symmetric neither with respect to the bright and the dark, nor with respect to the ${\lambda_z}/{4}$-shifts. This is clear with the quartet of the modes obtained in the previous section, which solutions can be denoted as $F,\tilde{F},F^*,$ and $\tilde{F}^*$ in the view of the (\ref{shifts}). Both pairs of the modes, $F,F^*$, and $\tilde{F},\tilde{F}^*$, are described by a pair of the wavenumbers $k_x,k_x^*$, and the symmetry (\ref{symmetry}) is broken.

The symmetry with respect to shifts by ${\lambda_z/4}$ (see the definition (\ref{shifts})), obviously, is not for the homogeneous medium uniquely. With full translation invariance in the $z$-direction broken, there can exist modulations that preserve that symmetry. An entire family of such is presented by a symmetric modification of the eq. (\ref{epsilon1}):
\begin{eqnarray}
\label{goff}
 \frac {{\lambda}_z^2 (k_0^2{\varepsilon} (z)-{k_x}^2)}{(2\pi)^2}=\frac{(4K(m))^2}{(2\pi)^2}\times\nonumber\\
 \left(1-\frac{m}{2}+
\frac{3}{4}\left(\sqrt{1-m}+1\right)^2\left(\left(\frac{{F}^2+{\tilde{F}}^2}{{F_0}^2}\right)^2-1\right)\right)\nonumber
\end{eqnarray}
\begin{figure}
  \includegraphics[width=3 in]{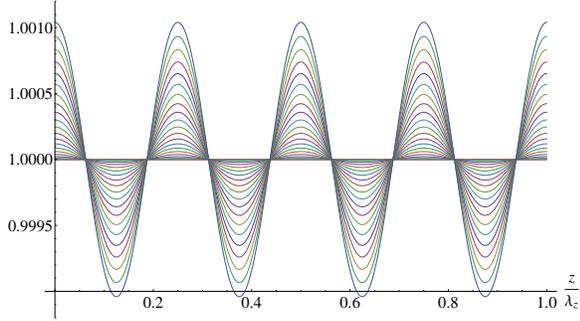}\\
  \caption{Symmetric modulations: $0\leq m \leq 0.1$.}\label{symm}
\end{figure}
The respective profiles of the permittivity modulation are clear from the \fref{symm}: plotted is the right-hand side of the above equation. The problem is equivalent to the so-called associated Lam\'{e} potentials \cite{Khare}.
In this symmetric case, the two modes have the same $k_x$: no Bragg birefringence is present. The modes, explicitly, are:
\begin{eqnarray}
\label{symmsANDc}
{F(z)=\frac{{F_0}(1-m)^{1/4}{{sn}\left(\left.u\right|m\right)}}{\sqrt{{dn}\left(\left.u\right|m\right)}}, \tilde{F}(z)=\frac{{F_0}{{cn}\left(\left.u\right|m\right)}}{\sqrt{{dn}\left(\left.u\right|m\right)}}}\nonumber
\end{eqnarray}
with $u\equiv\frac{4 K(m) z}{\lambda_z }$. The Wronskian $W(F,\tilde{F})$ (defined as ${\tilde{F}}{F'}-{F}{\tilde{F}'}$) is constant, because of the "degeneracy" mentioned above. This time the symmetry with respect to the inversion of the brightness does not hold, in contrast with the model presented by the eq. (\ref{epsilon1}): the respective mode $F^*(z)$ such that $(F^*(z))^2=F_0^2-(F(z))^2$ does not exist as an eigenmode for the given modulation of the permittivity; this is quite natural in the view of the degeneracy present in the second-order equation.

Gratings of this type can be of interest for the optically nonlinear photorefractive materials in the setups appropriate for the higher harmonic generation.

\section{Concluding remarks}\label{conclude}
In the view of the variety of possible mechanisms of the holographic grating formation in photorefractive media, it should be noted that even in a single material there may exist different mechanisms, due to the variations in the irradiation techniques (e.g., with pulsed irradiation, as compared to the cw irradiation, by varying the pulse duration and the intervals in between vs electronic relaxation times). It means that engineering the models is also possible.

In the case the holographic grating is recorded in a plate with $k_x$-controlled setup shown in the \fref{setups}, its thickness $l$ comes into play: because of reflections, it must match the phase of the interference pattern such that $2l=n\lambda_z$.
Similarly, in the case of using a mirror for recording, such considerations could help for better tuning the process, by slight variations of the angle of incidence and, if possible, the frequency of the incident beam.

In conclusion, it was discussed how the permittivity locally-isotropic modulation-induced anisotropy appears, leading to the Bragg birefringence effect, in the process of the recording of the holographic grating. The effect can be of interest for quantum devices, due to its phase-sensitivity.

\section*{Acknowledgments}

The work was supported by ISTC grant, Project A-1517.

\end{document}